# Title: Nonradiative DKR processes: revisiting the theory. I. Unimode model

Authors: Fabio DeMatteis (1) and Mladen Georgiev (2) (Dipartimento di Fisica, II Universita degli Studi di Roma "Tor Vergata", 00133 Roma, Italia (1) and Institute of Solid State Physics, Bulgarian Academy of Sciences, 1784 Sofia, Bulgaria (2))

We are revisiting the efficiency problem of DKR non-radiative transitions at local electronic centers in polarizable solids. Following a brief survey, we formulate the general rate equations thereby defining the microscopic parameters controlling the nonradiative deexcitation. We also reproduce quantum-mechanical expressions for the energy-conserving 'horizontal-transfer' reaction rates to compare with 'vertical' rates at various vibronic energy levels.

## 1. Introduction

### 1.1. Intrinsic quenching of the luminescence

Bartram and Stoneham [1,2] showed that the absence or presence of F-center luminescence after optical excitation follows a general criterion for the nonradiative deexcitation (DKR rule) first formulated by Dexter, Klick, and Russell [3]. The prediction of the luminescence quantum efficiency was formulated on the basis of parameters inferred from optical absorption data [4]. According to this (crossover) model, the nonradiative deexcitation occurs during lattice relaxation (dynamic process). A very small, but not vanishing probability for radiative transition is expected in NaBr and NaI. Although the small emission efficiency predicted by the crossover model was experimentally confirmed by Baldacchini, Pan, and Luty [5], these authors argued on the basis of the F-F' conversion efficiency and the temperature dependence of the emission quantum efficiency, that all excited F centers reach their relaxed excited states (RES). They explained the luminescence quenching by a subsequent 'vibronic tunneling' to the ground state from the excited electronic state in thermal equilibrium with the phonon bath (static process). A later work [6] tried to interpret the F-center emission efficiency and its wavelength dependence in several alkali halides according to the crossover theory. Their conclusions do not seem fully proved by the data shown therein.

### 1.2. Theory of dynamic processes

The precursor of any dynamic radiationless transition theory is the semiclassical Landau-Zener theory. (For details see, e.g. Bartram [8].) The electronic motion is treated quantum-mechanically by means of an interstate-coupling matrix element $V_{12}$, while the nuclear motion is described by a classical oscillator Q. The two potential energy curves $U_k(Q)$ are adiabatically split at the crossing point X corresponding to $Q_0$. The model predicts a maximum probability for the radiationless transition of 0.5 after one oscillation. In order to explain the extremely low quantum efficiency of luminescence, it is necessary to consider in detail the dissipative processes in the nuclear motion. If the damping of the oscillation in the excited state is slow enough, then multiple passages at $Q = Q_0$ will occur.

An approach to the crossover relaxation combining L-Z theory and the vibrational cooling is contained in the semiclassical treatment proposed by Sumi [8] which considers the classical trajectory of a vibrational wave-packet determined by a given set of vibrational modes.

Bartram and Stoneham have given a description of the opposite case in which a single sharp resonance dominates the spectrum of modes coupled to the electron center [2,4]. Assuming a coherent ultra-short (white) excitation pulse, they showed that the vibrational excitation in the (electronic} excited state can be associated with an effective local accepting mode. After the optical excitation, a non-stationary localized vibrational state will evolve and the local excitation will be gradually transferred to the lattice modes (vibrational cooling). The derivation relies on the requirement of well-separated vibrational levels for the effective local mode. If that inequality is not met, the excited state can relax by multiphonon transitions in a fraction of the vibrational period.

Leung and Song [9] have observed that the electronic transition is effective also at vibrational levels different from the one coinciding with the intersection of the potential curves. For each vibronic level in the excited electronic state, they consider two competing nonradiative processes: intralevel transition (pure vibrational relaxation) and interlevel transition (transition to a vibronic level in the ground electronic state). They have calculated the expected emission efficiency in several host lattices. Their calculations confirmed the prediction of the DKR model.

## 1.3. Theory of static processes

If the system is able to complete the relaxation to the RES (i.e. avoid a transition at the crossing point), then a thermal equilibrium with the phonon bath will be established. One can discuss its behavior in the framework of static processes, which include radiative transitions, thermal ionization and vibronic tunneling from the RES.

A general expression for the horizontal vibronic tunneling from a vibrationally thermalized excited state can be obtained on the basis of a theory that assumes linear electron-phonon coupling and a large equilibrium displacement in the excited state (strong coupling) with respect to the ground state [10].

## 1.4. Earlier investigations of the excited state lifetime in NaBr and NaI

The first experimental investigation of the F-center luminescence in NaBr was reported by Bosi et al. [11] which measured a weak photoconductivity signal and a faint luminescence signal at energies above 1 eV (cut-off energy of the photo-multiplier) [12]. The temporal profile of the decay of the luminescence was interpreted in terms of a dominant quick component (t = 23 ns) due to M centers and a slower (~μs) weak F-center contribution.

Although the measurements were apparently confirming the expectation of an extremely small efficiency for the optical emission, strong doubts on the interpretation have been cast by later accurate measurements of two different emission signals at 1.2 μm and 2.0 μm [5]. The 2.0 μm emission was shown to correspond to the optical relaxation from the RES, while the signal at higher energy was proposed to occur from a higher-lying level populated by recapture processes from the conduction band. Thus the 1.2 μm luminescence was related to the presence of F' centers. The emission quantum efficiency for the F band was estimated in the order of 0.2% by comparison with the integrated emission of the F center in a KBr crystal, which is known to have a quantum efficiency very close to 1. The temperature dependence of the F-F' conversion efficiency $\eta(T)$ was carefully investigated too. Assuming an Arrhenius-type ionization rate and a constant deexcitation rate, a decay time of 330 ps was estimated from the fit of the conversion efficiency data. The back conversion (F'-F) was found to occur with a full quantum efficiency under F'-band excitation. Similar results were obtained in NaI samples. Two emission bands were reported at 1.65 μm and 2.2 μm for F' and F centers, respectively. The emission quantum efficiency for the F band was estimated at 0.5% and the deexcitation time about 10 ps.

Beside the direct optical excitation in the F band, the F-luminescence is generated also by the recapture of a conduction electron at the vacancy into the RES following the F' optical excitation [13]. Baldacchini et al. [5] argued that the nonradiative transition responsible for the quenching of the F luminescence must occur from the RES, since this recapture luminescence had the same low quantum efficiency as that excited by the direct optical excitation.

### 1.5. Further experimental developments

Schoemaker and co-workers have contributed substantially to the topic by studying relaxation and ionization processes at the F center in NaBr [14] and NaI [15] with picosecond laser pulses. In NaBr they found that a nonradiative horizontal tunneling deexcitation from a thermalized RES can also account for the observed meagre emission efficiency as does the DKR criterion. In NaI the contribution of the non-equilibrium DKR de-excitation during the vibrational relaxation seems to be evidenced by the observed correlation of the time constant for ground-state recovery with the vibrational lifetime of the resonant Raman spectrum from line-width measurements. Although this does not solve the controversy over the DKR de-excitation, it stresses the necessity of further theoretical studies.

## 2. De-excitation rate equations

Referring to Fig. 1 for an introduction to the quantities involved in the subsequent discussion, we consider a local phonon-coupled electronic center with a ground electronic state $F_g$ and an excited electronic state $F_e$, the latter of which becomes occupied following an optical excitation. Let $E_n$ and $N_n$ be the energies and populations, respectively, of the vibronic levels in $F_e$. The occupation $N_n$ of any individual level $E_n$ in $F_e$ depleting due to two de-excitation channels: a horizontal tunneling rate $k_{hn}$ at constant energy towards $F_g$ and a vertical tunneling rate $k_{vn}$ down to $E_{n-1}$, and enhancing due to $k_{vn+1}$ from $E_{n+1}$, the following rate equation holds good:

$$(d/dt)N_n = -(k_{hn} + k_{vn})N_n + k_{vn+1}N_{n+1} \quad (1)$$

We consider a nearest-neighbor coupling and neglect higher-order terms of the form $k_{vn+k}N_{n+k}$, etc.

The total level occupancy at t = 0 is

$$\sum_n N_n(0) = N(0)\ (\equiv c_F\, t_p I), \quad (2)$$

$c_F$ being the photon absorption cross section, $t_p$ and I the light-pulse duration and intensity, respectively. Differentiating (2) on using (1) one gets (the 0-indexed brackets dropped)

$$0 = -\sum_n(k_{hn} + k_{vn})N_n + \sum_n k_{vn+1}N_{n+1}, \quad (3)$$

so that if $k_{vn} \equiv k_v$ = const then $\sum_n k_{hn}N_n + k_v N_0 = 0$; confirming $k_{vn} \approx$ const. $E_0$ being the lowest vibronic level, we set $k_{v0} = k_{rad}$, the radiative de-excitation rate from the relaxed excited state (RES).

If the ultimate occupied level is $E_m$ then $N_{m+1} = 0$ and

$$(d/dt)N_m = -(k_{hm} + k_{vm})N_m, \quad (4)$$

which is readily solved to give

$$N_m(t) = N_m(0)\,exp(-[k_{hm}+k_{vm}]t) \quad (5)$$

The subsequent equation for m-1 being

$$(d/dt)N_{m-1} = -(k_{hm-1} + k_{vm-1})N_{m-1} + k_{vm}N_m(t)$$

$$= -(k_{hm-1} + k_{vm-1})N_{m-1} + k_{vm}N_m(0)\,exp(-[k_{hm}+k_{vm}]t)$$

it gives

$$N_{m-1}(t) = exp(-[k_{hm-1}+k_{vm-1}]t)\{N_{m-1}(0)$$

$$+ \ {}_0\!\int^t dt' k_{vm}N_m(0)\,exp(-[k_{hm}+k_{vm}]t')\,exp([k_{hm-1}+k_{vm-1}]t')\}$$

$$= exp(-[k_{hm-1}+k_{vm-1}]t)\{N_{m-1}(0)$$

$$+ N_m(0)k_{vm}(k_{hm-1}+k_{vm-1}-k_{hm}-k_{vm})^{-1}\ [exp([k_{hm-1}+k_{vm-1}-k_{hm}-k_{vm}]t) - 1]\}$$

$$= N_{m-1}(0)\,exp(-[k_{hm-1}+k_{vm-1}]t) + N_m(0)k_{vm}(k_{hm-1}+k_{vm-1}-k_{hm}-k_{vm})^{-1}$$

$$\times\ [exp(-[k_{hm}+k_{vm}]t) - exp(-[k_{hm-1}+k_{vm-1}]t)]$$

Further, we have

$$(d/dt)N_{m-2} = -(k_{hm-2}+k_{vm-2})N_{m-2} + k_{vm-1}N_{m-1}(t)$$

$$= -(k_{hm-2}+k_{vm-2})N_{m-2} + k_{vm-1}\{N_{m-1}(0)exp(-[k_{hm-1}+k_{vm-1}]t)$$

$$+ N_m(0)k_{vm}(k_{hm-1}+k_{vm-1}-k_{hm}-k_{vm})^{-1}\times[exp(-[k_{hm}+k_{vm}]t) - exp(-[k_{hm-1}+k_{vm-1}]t)]\}$$

leading to

$$N_{m-2}(t) = N_{m-2}(0)exp(-[k_{hm-2}+k_{vm-2}]t) + N_{m-1}(0)k_{vm-1}[k_{hm-2}+k_{vm-2}-k_{hm-1}-k_{vm-1}]^{-1}$$

$$\times [exp(-[k_{hm-1}+k_{vm-1}]t) - exp(-[k_{hm-2}+k_{vm-2}]t)]$$

$$+ N_m(0)k_{vm-1}k_{vm}(k_{hm-1}+k_{vm-1}-k_{hm}-k_{vm})^{-1}\{(k_{hm-2}+k_{vm-2}-k_{hm}-k_{vm})^{-1}exp(-[k_{hm}+k_{vm}]t)$$

$$- (k_{hm-2}+k_{vm-2}-k_{hm-1}-k_{vm-1})^{-1}exp(-[k_{hm-1}+k_{vm-1}]t)\}$$

Inasmuch as the rate equation for m-k is

$$(d/dt)N_{m-k} = -(k_{hm-k} + k_{vm-k})N_{m-k} + k_{vm-k+1}N_{m-k+1}(t) \quad (6)$$

we generalize the solution to read

$$N_{m-k}(t) = N_{m-k}(0)exp(-[k_{hm-k}+k_{vm-k}]t) + N_{m-k+1}(0)k_{vm-k+1}[k_{hm-k}+k_{vm-k}-k_{hm-k+1}-k_{vm-k+1}]^{-1}$$

$$\times [exp(-[k_{hm-k+1}+k_{vm-k+1}]t)-exp(-[k_{hm-k}+k_{vm-k}]t)] + ...$$

$$+ N_m(0)k_{vm-1}k_{vm}(k_{hm-1}+k_{vm-1}-k_{hm}-k_{vm})^{-1}\{(k_{hm-k}+k_{vm-k}-k_{hm}-k_{vm})^{-1}exp(-[k_{hm}+k_{vm}]t)$$

$$- (k_{hm-k}+k_{vm-k}-k_{hm-1}-k_{vm-1})^{-1}exp(-[k_{hm-1}+k_{vm-1}]t)\} \quad (7)$$

What is essential is that the solution in equation (7) is a sum of exponentials of different time slopes whose measurement can be utilized to determine the elemental horizontal and vertical transition rates.

### 2.2. Example: a three-level system

To illustrate the above general rate equations, we consider a three-level system which may be close to describing the actual situation under monochromatic laser excitation. The following rate equations apply:

$$(d/dt)N_2 = -(k_{h2} + k_{v2})N_2$$

$$(d/dt)N_1 = -(k_{h1} + k_{v1})N_1 + k_{v2}N_2$$

$$(d/dt)N_0 = -(k_{h0}+k_{rad})N_0 + k_{v1}N_1 \quad (8)$$

We subsequently solve for the populations to arrive at

$$N_2(t) = N_2(0)exp(-[k_{h2}+k_{v2}]t)$$

$$N_1(t) = N_1(0)exp(-[k_{h1}+k_{v1}]t) + N_2(0)k_{v2}(k_{h1}+k_{v1}-k_{h2}-k_{v2})^{-1}$$

$$\times \{exp(-[k_{h2}+k_{v2}]t) - exp(-[k_{h1}+k_{v1}]t)\}$$

$$= \{N_1(0)-N_2(0)k_{v2}(k_{h1}+k_{v1}-k_{h2}-k_{v2})^{-1}\}exp(-[k_{h1}+k_{v1}]t)$$

$$+ N_2(0)k_{v2}(k_{h1}+k_{v1}-k_{h2}-k_{v2})^{-1}exp(-[k_{h2}+k_{v2}]t)$$

$$N_0(t) = N_0(0)exp(-[k_{h0}+k_{rad}]t) + N_1(0)k_{v1}(k_{h0}+k_{rad}-k_{h1}-k_{v1})^{-1}$$

$$\times [exp(-[k_{h1}+k_{v1}]t) - exp(-[k_{h0}+k_{rad}]t)] + N_2(0)k_{v1}k_{v2}(k_{h1}+k_{v1}-k_{h2}-k_{v2})^{-1}$$

$$\times \{(k_{h0}+k_{rad}-k_{h2}-k_{v2})^{-1}exp(-[k_{h2}+k_{v2}]t) - (k_{h0}+k_{rad}-k_{h1}-k_{v1})^{-1}exp(-[k_{h1}+k_{v1}]t)\}$$

$$= \{N_0(0)-N_1(0)k_{v1}(k_{h0}+k_{rad}-k_{h1}-k_{v1})^{-1}\}exp(-[k_{h0}+k_{rad}]t)$$

$$+ \{N_1(0)k_{v1}(k_{h0}+k_{rad}-k_{h1}-k_{v1})^{-1}-N_2(0)k_{v1}k_{v2}(k_{h1}+k_{v1}-k_{h2}-k_{v2})^{-1}$$

$$\times (k_{h0}+k_{rad}-k_{h1}-k_{v1})^{-1}\}exp(-[k_{h1}+k_{v1}]t) + N_2(0)k_{v1}k_{v2}$$

$$\times (k_{h1}+k_{v1}-k_{h2}-k_{v2})^{-1}(k_{h0}+k_{rad}-k_{h2}-k_{v2})^{-1}exp(-[k_{h2}+k_{v2}]t) \qquad (9)$$

arranged in the form of a series of exponentials.

### 2.3. Transition rates

Depending on the relative values of $k_{hn}$ and $k_{vn}$, the DKR nonradiative de-excitation prevails for $k_{hn} \gg k_{vn}$, while for $k_{hn} \ll k_{vn}$ a thermal equilibrium is established in $F_e$ through the faster intra-vibrational relaxation. Both rates can be derived quantum-mechanically.

#### 2.3.1. Horizontal rate $k_{hn}$

At a given vibronic level $E_n$ in the electronic excited state $F_e$, the horizontal energy-conserving de-excitation transition occurs as configurational tunneling across a potential energy barrier to the corresponding vibronic level $E_m$ in the electronic ground state $F_g$. Following Christov [16], the transition rate $k_{hn}$ of this elemental process will be constructed as the product (Condon's approximation!) of the configurational tunneling probability $W_{conf}(E_n)$ times the probability $W_{el}(E_n)$ (not necessarily normalized) for changing the electronic state in the course of the de-excitation:

$$k_{hn} = \nu_{vib}W_{el}(E_n)W_{conf}(E_n) \qquad (10)$$

Here $\nu_{vib}$ is the vibrational frequency, assumed the same in both $F_g$ and $F_e$. If $V_{eg}$ is the splitting of the electronic energy $E_C$ at crossover, then our subsequent discussion holds good for $V_{eg} \ll E_C$ [14].

$W_{el}$ has been derived by means of Landau-Zener's method taking account for the possibility of a multiple passage of the system through the crossover point:

$$W_{el}(E_n) = 2[1 - exp(-2\pi\gamma(E_n))]exp(-2\pi\gamma(E_n)),\ E_n >> E_C$$

$$W_{el}(E_n) = 2\pi\gamma_n(1-2\gamma_n)exp(2\gamma_n)/\Gamma(1-\gamma_n),\ E_n << E_C \qquad (11)$$

Here

$$\gamma(E_n) = (V_{eg}^2/ 2\hbar\omega_{vib})[E_R\,|\,E_n\text{-}E_C\,|\,]^{-1/2} \qquad (12)$$

is Landau-Zener's parameter, $E_R$ is the reorganization energy.

The configurational-tunneling probability has been derived to read:

$$W_{conf}(E_n) = [(F_{mn}^s/F_{nn}^w)^2(2^n n!)/(2^m m!)]exp(2Q/ \hbar\omega_{vib}) \qquad (13)$$

where

$$F_{nn}^w = 2\xi_c H_n(\xi_c)H_n(-\xi_c) - 2nH_{n-1}(\xi_c)H_{n-1}(-\xi_c) + 2nH_n(\xi_c)H_{n-1}(-\xi_c)$$

$$F_{nm}^s = \xi_m H_n(\xi_c)H_m(\xi_c-\xi_m) - 2nH_{n-1}(\xi_c)H_m(\xi_c-\xi_m) + 2mH_n(\xi_c)H_{m-1}(\xi_c-\xi_m)$$

with $H_n(\xi)$ standing for Hermite's polynomials of n-th order, $\xi=(M\omega_{vib}^2/ \hbar\omega_{vib})^{1/2} q$ is the dimensionless mode coordinate; $\xi_g$ and $\xi_e$ stand for the extremal configurational coordinates in $F_g$ and $F_e$, respectively, $\xi_c$ is the crossover coordinate. $n \equiv n_i$ and $m \equiv n_f$ are the vibrational quantum numbers in the initial and final electronic states, respectively.

It will be instructive to study the behavior of $W_{conf}(E_n)$ at large $Q/\hbar\omega_{vib} = m\text{-}n$ using Sterling's formula for the factorials $m! \sim m^m exp(-m)$, etc. Inserting into (13) we get for $m \gg n$:

$$W_{conf}(E_n) \sim \{\xi_{2m} / [2^m m^m exp(-m)]\}exp(2[m\text{-}n])$$

which gives $W_{conf} \ll 1$ for $ln(m) \gg 3\text{-}ln(2)+2ln(\xi)$ typically leading to $ln(m) \gg 3.7$, $m \gg 40$. At large $Q/ \hbar\omega_{vib} \sim 100$, therefore, $W_{conf}$ will be vanishingly small. Note that large Q's are characteristic of most F centers in alkali halides.

The overall horizontal rate is next obtained by summing up all elemental rates weighed by the occupation probabilities:

$$k_{eg} = \Sigma_n(N_n/N)k_{hn} \qquad (14)$$

If thermal equilibrium is rapidly established on $k_{vn} \gg k_{hn}$, then

$N_i / N = (1/Z)\exp(-E_n/k_BT) = 2\sinh(\hbar\omega_{vib} / 2k_BT)\exp(-E_n/k_BT)$ (15)

and the overall horizontal rate becomes

$k_{eg}(T) = 2\sinh(\hbar\omega_{vib}/2k_BT)\sum_n k_{hn} \exp(-E_n/k_BT)$

$= 2\sinh(\hbar\omega_{vib}/2k_BT)\nu_{vib}\sum_n W_{el}(E_n)W_{conf}(E_n)\exp(-E_n/k_BT)$ (16)

for deexcitation from the relaxed excited state.

2.3.2. Vertical rate $k_{vn}$

The vertical tunneling rate $k_{vn}$ is controlled by the inter-vibrational coupling. If $\Delta\omega_n$ is the width of the inter-vibrational frequency splitting (frequency bandwidth) at $E_n$ then ($\Delta\omega_n = \Delta\omega$) [2]:

$k_{vn} = n\pi\Delta\omega$ (17)

The analytic expression relates $\Delta\omega$ to parameters inherent to the crystal, and indeed, for a weak frequency dispersion of the form

$\omega_k = (\omega_0^2 + \omega_1^2\cos k)^{½} \sim \omega_0 + (\omega_1^2/ 2\omega_0)\cos k$

the vibrational bandwidth is [17]:

$\Delta\omega = \omega_1^2/ 2\omega_0$.

With $½M\omega_1^2 q_i q_j$ standing for the intra-vibrational coupling term in a small-polaron Hamiltonian, $\omega_{12}$ is a coupling frequency constant. Numerically, $\pi\Delta\omega \sim 5\times10^{10}$ has been used [9].

2.3.3. Configurational coordinate model

To illustrate the comparison of the two elemental de-excitation rates at $E_n$, we consider a simple electron & mode Hamiltonian at an F center:

$H = H_e + H_{mode} + H_{int}$ (18)

$H_e = \mathbf{p}^2/ 2m + V(r,0)$

$H_{mode} = \mathbf{P}^2/ 2M + ½Kq^2$

$H_{int} = b(r)q \equiv (\partial/\partial q)V(r,q)\big|_{q=0}\, q$

V(r,q) is the potential acting upon the F center electron which is modulated by the mode coordinate q. Linear electron-mode coupling is effected through the first-order term of the series expansion of V(r,q) in q. Following standard procedures explained elsewhere [18], a static electronic basis has been constructed by choosing two orthonormal eigenstates $|g>$ and $|e>$ of $H_e$. We get the following 'adiabatic' eigenvalues:

$$E_{\pm}(q) = ½\{V_g(q)+V_e(q) \pm [(V_e(q)-V_g(q))^2 + 4V_{eg}V_{ge}]^{½}\} \qquad (19)$$

where $V_g(q)$ and $V_e(q)$ are the 'diabatic' parabolae:

$$V_g(q) = ½Kq^2 + <g|b(r)|g>q + <g|H_e|g> \equiv ½K(q-q_g)^2 + Q_g$$

$$V_e(q) = ½Kq^2 + <e|b(r)|e>q + <e|H_e|e> \equiv ½K(q-q_e)^2 + Q_e$$

with

$$q_g = -<g|b(r)|g>/K$$

$$q_e = -<e|b(r)|e>/K$$

$$Q_g = -½Kq_g^2 + <g|H_e|g>$$

$$Q_e = -½Kq_e^2 + <e|H_e|e>$$

$$Q = Q_e - Q_g$$

whereby we calculate

$$E_R = ½K(q_e-q_g)^2$$

$$q_C = (q_e^2 - q_g^2 + 2Q/K)/2(q_e-q_g)$$

$$E_C = ½K(q_c-q_e)^2$$

$$V_{eg} = <e|b(r)|g>q_c, \; V_{ge} = <g|b(r)|e>q_c$$

To make specific calculations on the vibronic *F* center, we use a simple semi-continuum electronic potential $V(r,0)$ composed of a spherical well at $r \le r_0$ to represent the anion vacancy and a Coulomb tail at $r \ge r_0$ to describe the longer-range attraction, where $r_0$ is the well radius. The bound eigenstates of $H_e$ are $\Psi(r) = \psi(r)Y(\varphi,\theta)$ where $\psi(r)$ are spherical Bessel functions at short range and hydrogen-like wave-functions at long range [19]:

$$\psi(r) = A_{\kappa l}j_l(\kappa r), \; (r \le r_0) = B_{\alpha l}R_{nl}(\alpha r), \; (r \ge r_0)$$

We modulate that potential by the mode coordinate q:

$$V(r,q) = -V_0(r_0,q)[1-\Theta(r-r_0-q)] - (e^2/\varepsilon r)\Theta(r-r_0-q),$$

where $\Theta(r)$ is the step function and $V_0(r,q) \sim 1/(r_0+q)$ is the well depth, so as to effect an electron-mode coupling of the form $b(r) \equiv \partial V(r,q)/\partial q\,|_{q=0} = [V_0(r_0,0)/r_0][1-\Theta(r-r_0)] - [V_0(r_0,0) - (e^2/\varepsilon r)]\delta(r-r_0)$

with matrix elements

$< \psi_{l\kappa\alpha}\,|\,b(r)\,|\,\psi_{l'\kappa'\alpha'} > = [V_0(r_0,0)/r_0] <j_{\kappa l}\,|\,j_{\kappa' l'}> - [V_0(r_0,0) - (e^2/\varepsilon r_0)]\psi_{l\kappa\alpha}(r_0)\psi_{l'\kappa'\alpha'}(r_0)$

Details can be found in Ref. [18] and [19] where semi-continuum-potential quantities are calculated for NaI and some of these are now reproduced in Table I (cf. also Ref. [20]-[22]). Three states are considered, 1s- and 2s- (l=0) and 2p- like (l=1), as follows:

$\psi_{1s}(r) = A_{1s}(\kappa r)^{-1}\sin(\kappa r)$ (in), $B_{1s}\exp(-\alpha r)$ (out)

$\psi_{2s}(r) = A_{2s}(\kappa r)^{-1}\sin(\kappa r)$ (in), $B_{2s}(1-\alpha r)\exp(-\alpha r)$ (out)

$\psi_{2p}(r) = A_{2p}(\kappa r)^{-1}\,[\sin(\kappa r)/(\kappa r)-\cos(\kappa r)]$ (in), $B_{2p}\alpha r\exp(-\alpha r)$(out)

etc. with $\kappa$ and $\alpha$ differing for the different states. Although archaic the semi-continuum model has proved instrumental in revealing the essential physics. The diabatic potentials generated by the three eigenstates are shown in Figure 1, calculated using Table I data.

### 2.3.4. Horizontal versus vertical rate

To compare between $k_{hn}$ and $k_{vn}$, considering overbarrier levels would suffice, in so far as checking whether $k_{hn} \ll k_{vn}$ is the main concern, while $W_{conf}(E_n) \sim 1$ at $E_n \gg E_B$. From equation (11) we have for $E_n \gg E_C$:

$k_{hn} / k_{vn} \leq 2\nu_{vib}[1-\exp(-2\pi\gamma(E_n))]\exp(-2\pi\gamma(E_n))/n\pi\Delta\omega \leq 4\pi\nu_{vib}\gamma(E_n)/n\pi\Delta\omega$,

the latter inequality holding good for $2\pi\gamma(E_n) \ll 1$. For an overbarrier level, $E_n \geq E_C \sim E_B$ implying $n \geq E_C / \hbar\omega - ½ \sim E_C / \hbar\omega$, we get

$k_{hn}/k_{vn} \leq 4\nu_{vib}[\gamma(E_n)/E_C](\hbar\omega/\Delta\omega) = (\omega/\pi\Delta\omega)(V_{eg}^2/E_C\,[E_R\,|\,E_n-E_C\,|\,]^{½})$

$= (1/\pi[SP_n]^{½})(V_{eg}/E_C)(V_{eg}/\,\hbar\Delta\omega) \ll V_{eg} / \hbar\Delta\omega$

setting $P_n = (E_n - E_C) / \hbar\omega$ and $S = E_R / \hbar\omega$, the Huang-Rhys factor. In as much as typically S, $P_n \sim 10$ and in view of $V_{eg} / E_C \ll 1$, the ratio $k_{hn} / k_{vn} \ll 1$ unless $V_{eg} / \hbar\Delta\omega \gg 1$. We see that the DKR efficiency is crucially dependent on the magnitude of the ratio of the static electronic splitting to the intra-vibrational splitting at the local electronic center [cf. Ref. [23]}.

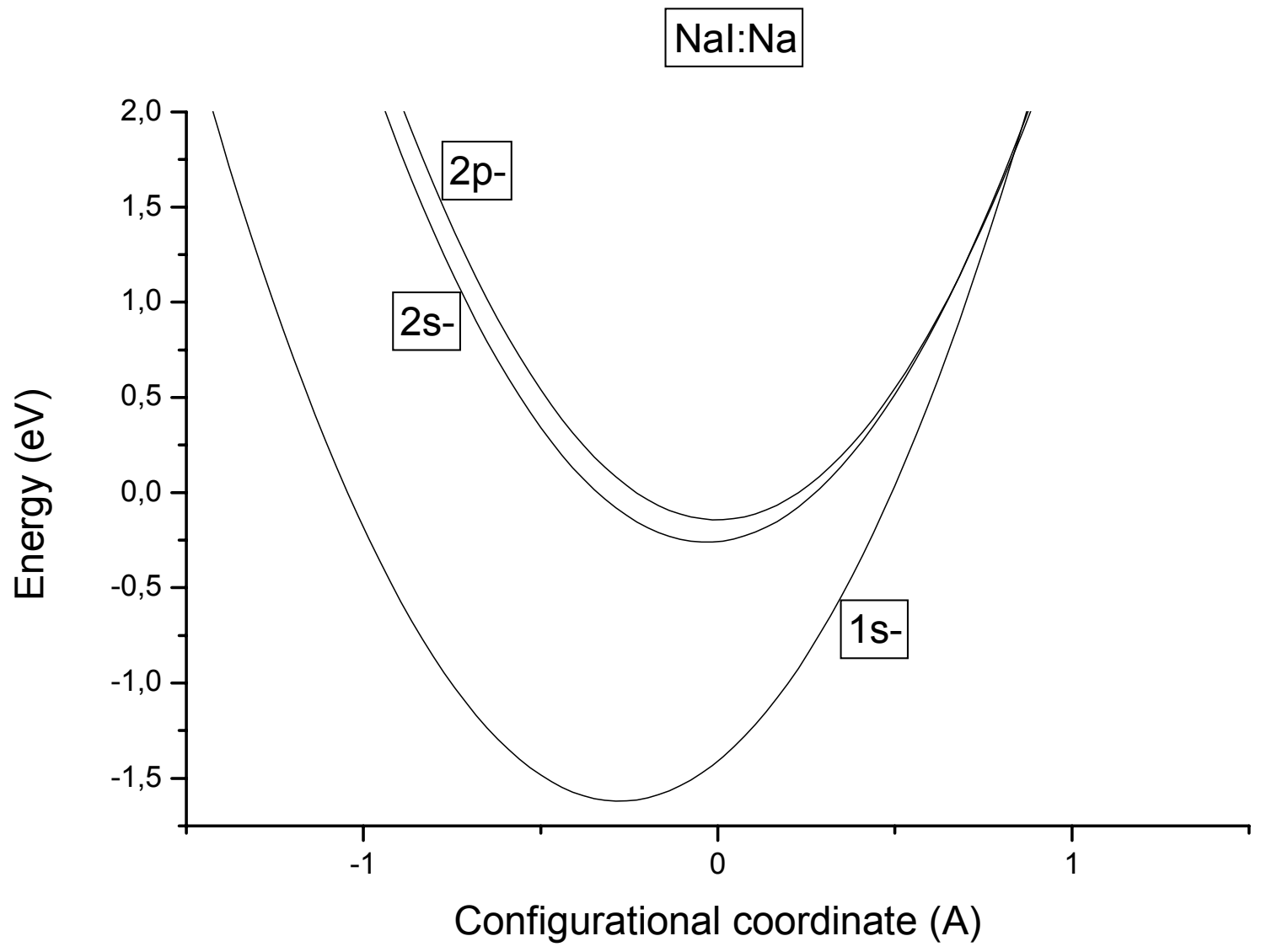

Figure 1: Diabatic potentials of the NaI F center in the harmonic approximation with a breathing-mode coupling, as calculated using the parameters in Table I. The parabolae pertain correspondingly to the 1s-, 2s-, and 2p- like F center states. The particular configurational-coordinate diagram does not suggest any dynamic crossover processes to quench the F center luminescence except for ones incited at higher photon energies.

It is usually considered that the horizontal rate $k_{hn}$ may only be competitive to the vertical rate $k_{vn}$ if $E_n$ is an over-barrier level. This requirement can be given a simple mathematical form, namely: $E_n \sim (n + ½)\ \hbar\omega_{vib}+Q \geq E_B \sim E_C - V_{eg} = (E_R+Q)^2/4E_R - V_{eg}$ which leads to $\Lambda \equiv S\hbar\omega_{vib} / E_{max} \leq ¼$ for $E_B \gg V_{eg}$ and $E_n \equiv E_{max} = E_R+Q$, the peak energy reached in absorption; here $E_R = S\hbar\omega_{vib}$, S being the Huang-Rhys factor. $\Lambda$ therefore is the ratio of the relaxed phonon energy to the absorbed optical energy. $\Lambda < ¼$ is the criterion that the system may reach RES and emit luminescent light following the optical absorption.

However it should be stressed immediately that the over-barrier term in (11) is vanishing at both large and small $\gamma$'s and maximalizes to 0.5 at $2\pi\gamma(E_n)= \ln 2$. This peculiar behavior makes the over-barrier electron-transfer virtually non-competitive to the vibrational relaxation at vibronic energy levels both sub-barrier near the barrier top (large $\gamma$'s) and over-barrier far above the top (small $\gamma$'s). The configurational rearrangement occuring with near certainty above the barrier, we see that horizontal de-excitation processes may virtually be expected competitive only at over-barrier energy levels around

$$E_n \sim E_C + 20.54(V_{eg}^2/ \hbar\omega_{vib})^2/E_R = E_C + 20.54\ \hbar\omega_{vib}(V_{eg} / \hbar\omega_{vib})^4/S$$

where $S = E_R / \hbar\omega_{vib}$ is the Huang-Rhys factor again. It is informative to check just how these $E_n$ compare with the peak energy $E_R$ in absorption. We have

$$E_n / E_R = ¼(1+Q/E_R)^2 + 20.54(V_{eg} / \hbar\omega_{vib})^4/S^2.$$

From Table I we have $S \sim 20$ (at $E_R \sim 0.2$ eV and $\hbar\omega_{vib} \sim 0.01$ eV) and also $V_{eg} \sim 0.1$ eV, $Q \sim -1.4$ eV, all for the 2p-1s transitions, yielding $E_n / E_R \sim 500 >> 1$ coming mainly from the second term. We conclude that DKR de-excitations may eventually be observed at the far short-wavelength side of the F band.

### 2.4. Equilibrium non-radiative de-excitation rate

At $k_{vn} \gg k_{hn}$, an overall horizontal de-excitation occurs from the thermalized vibronic energy levels in $F_e$, a situation usually referred to as Relaxed Excited State $F_e \sim$ (RES). The corresponding rate being given by equation (16), it has a specific temperature dependence exhibiting a finite zero-point rate followed gradually by an Arrhenius-like sloped portion at the higher temperatures [16].

### 2.5. Luminescent quantum yield

The luminescent quantum yield $\eta$ from the lowest vibronic level in a non-thermalized electronic excited state $F_e$, as the system undergoes either vertical relaxation or horizontal de-excitation steps, has been derived by Leung and Song [9] to be:

$$\eta = \prod_n [k_{vn}/(k_{vn}+ k_{hn})] \qquad (20)$$

where $k_{v0} = 0$. Equation (20) yields $\eta = 1$ for $k_{hn} \ll k_{vn}$. However, if we set as above $k_{v0} = k_{rad}$ and let thermalization go, then the yield for luminescence from *RES* becomes

$$\eta(T) = k_{rad} / (k_{rad}+k_{nonrad}) \qquad (21)$$

where $k_{nonrad} = k_{eg}$ is again given by equation (16).

## 3. Conclusion

The foregoing simple theory can serve as a basis for more sophisticated studies of the DKR efficiency at local centers in specific host crystals. It sets forth the frame for defining the horizontal de-excitation rates in terms of the quantum-mechanical reaction-rate theory.

However, the present model being based on a single vibrational frequency, it does not adequately account for the crossover barriers. In NaI, both crossover barriers are calculated too high (~ 2 eV) based on the $A_{1g}$ mode vibrational frequency obtained in Raman experiments, as explained elsewhere.[21,22]

Further extensions should be made redefining the theory for more realistic situations where two different vibrational frequencies couple to the ground and excited electronic states, respectively. Only then will detailed calculations be made and compared with experimental data on F centers in a variety of crystalline hosts.

Table I

Selfconsistent Semicontinuum-Potential Calculations for NaI
(Breathing-Mode Coupling)

| State | Cavity Radius $r_0$ (Å) | Cavity Potential $V_0$ (eV) | Dielectric Constant $\varepsilon$ | Wavefunction Parameters $u=\alpha r_0$ | $v=\kappa r_0$ | Normalized Constants A ($Å^{-3/2}$) | B ($Å^{-3/2}$) |
|---|---|---|---|---|---|---|---|
| 1s- | 3.237 | 5.003 | 3.100 | .985 | 1.562 | .201 | .345 |
| 2s- | 3.237 | 4.924 | 2.965 | .515 | 1.870 | .072 | .127 |
| 2p- | 3.237 | 4.557 | 3.510 | .435 | 1.430 | .042 | .057 |

| State | Electron Energy $<\psi\|H_e\|\psi>$ (eV) | Coupling Constant $<\psi\|b(r)\|\psi>$ (eV/Å) | Vibrational Quantum $\hbar\omega_{vib}$ (meV) | Stiffness $K=M\omega^2$ ($eV/Å^2$) | Configurational Coordinates (Å) $q_{min}$ | $q_{crossover}$ 1s- | $q_{crossover}$ 2s- |
|---|---|---|---|---|---|---|---|
| 1s- | -1.410 | 1.520 | 13 | 5.492 | -.277 | | .857 |
| 2s- | -0.257 | 0.174 | 13 | 5.492 | -.032 | .857 | |
| 2p- | -0.143 | 0.006 | 13 | 5.492 | -.001 | .837 | .679 |

| State | Lattice Relaxation Energy $E_R$ (eV) 1s- | Crossover Energy Barrier $E_C$ (eV) 1s- | $E_C$ (eV) 2s- | Electron Binding Energy $Q_\psi$ (eV) | Electron Energy Splitting $V_{eg}$ (eV) 1s- | $V_{eg}$ (eV) 2s- | Vibrational Splitting $\hbar\Delta\omega_{vib}$ (meV) |
|---|---|---|---|---|---|---|---|
| 1s- | | | 2.170 | -1.621 | | .104 | $10^{-2}$ |
| 2s- | 0.165 | 2.170 | | -0.260 | .103 | | $10^{-2}$ |
| 2p- | 0.209 | 1.928 | 1.270 | -0.143 | 0 | 0 | $10^{-2}$ |

## Appendix I

### Calculation of the electron-exchange matrix elements
### (The off-diagonal matrix elements)

The relevant radial integrals are:

$I_{ss} = \langle j_{0\kappa}(r) | j_{0\kappa'}(r)\rangle = AA'{}_0\int^{r0} drr^2(\kappa r)^{-1}(\kappa' r)^{-1} sin(\kappa r) sin(\kappa' r)$

$= (AA'r_0^3/vv')\ {}_0\int^1 dx\, sin(vx) sin(v'x) = (AA'r_0^3/2vv')\ {}_0\int^1 dx\{cos([v-v']x) - cos([v+v']x)\}$

$= (AA'r_0^3/2vv')\{(v-v')^{-1} sin(v-v') - (v+v')^{-1} sin(v+v')\}$

$I_{sp} = \langle j_{0\kappa}(r) | j_{1\kappa'}(r)\rangle = AA'{}_0\int^{r0} drr^2(\kappa r)^{-1} sin(\kappa r)(\kappa' r)^{-1}\{sin(\kappa' r)/(\kappa' r) - cos(\kappa' r)\}$

$= (AA'r_0^3/vv')\ {}_0\int^1 dx\, sin(vx)\{sin(v'x)/(v'x) - cos(v'x)\}$

$= (AA'r_0^3/vv')\ {}_0\int^1 dx\{sin(vx) sin(v'x)/(v'x) - sin(vx) cos(v'x)\}$

$= (AA'r_0^3/2vv')\ {}_0\int^1 dx\{[cos([v-v']x) - cos([v+v']x)]/(v'x)$

$- [sin([v+v']x) + sin([v-v']x)]\}$

$= (AA'r_0^3/2vv')\{[ci(v-v') - ci(v+v')]/v' - 2v/(v+v')(v-v')$

$+ (v+v')^{-1} cos(v+v') + (v-v')^{-1} cos(v-v')\}$

where $ci(x) = -{}_x\int^{\infty} dt\, cos(t)/t$ is the integrated cosine function which is tabulated [24].

Using $I_{ss}$ and $I_{sp}$, we now calculate radial off-diagonal matrix elements following the prescription:

$V_{\psi\psi'} = \{[V_0(r_0,0)/r_0]\langle j_{\kappa l} | j_{\kappa' l'}\rangle - [V_0(r_0,0) - (e^2/\varepsilon r_0)]\ j_{\kappa l}(r_0)\ j_{\kappa' l'}(r_0)\} q_C$

in terms of spherical Bessel functions only:

$V_{1s\text{-}2s} = AA'q_C\{[V_0(r_0,0)/r_0](r_0^3/2vv')[(v-v')^{-1} sin(v-v')$

$- (v+v')^{-1} sin(v+v')] - [V_0(r_0,0) - (e^2/\varepsilon r_0)][sin(v)/v][sin(v')/v']\}$

$V_{1s\text{-}2p} = AA'q_C\{[V_0(r_0,0)/r_0](r_0^3/2vv')\{[ci(v-v')-ci(v+v')]/v'$

$- 2v/(v+v')(v-v') + (v+v')^{-1} cos(v+v')+(v-v')^{-1} cos(v-v')\}$

$- [V_0(r_0,0) - (e^2/\varepsilon r_0)] sin(v)[sin(v')/(v')-cos(v')]\}$

$V_{2s\text{-}2p} = AA'q_C\{[V_0(r_0,0)/r_0](r_0^3/2vv')\{[ci(v-v') - ci(v+v')]/v'$

$- 2v/(v+v')(v-v') + (v+v')^{-1} cos(v+v') + (v-v')^{-1} cos(v-v')\}$

$- [V_0(r_0,0) - (e^2/\varepsilon r_0)]\mathit{sin}(v)[\sin(v')/(v') - \mathit{cos}(v')]\}$

where $v = \kappa r_0$, etc.

The radial matrix elements should be corrected for the angular parts of the square-wave wave functions $Y_{lm}(\theta,\varphi)$, namely $Y_{00}(\theta,\varphi) = 1/\sqrt{(4\pi)}$, $Y_{10}(\theta,\varphi) = \sqrt{(3/4\pi)}\cos\theta$ leading to:

$$<j_{0K}Y_{00}|Y_{00}j_{0K'}> = <Y_{00}|Y_{00}><j_{0\kappa}|j_{0\kappa'}> = <j_{0\kappa}|j_{0\kappa'}>$$

$$<j_{0\kappa}Y_{00}|Y_{10}j_{1\kappa'}> = <Y_{00}|Y_{10}><j_{0\kappa}|j_{1\kappa'}> = 0,$$

etc. This gives the complete matrix element in the form:

$$V_{\psi\psi'} = \{[V_0(r_0,0)/r_0]<Y_{lm}|Y_{l'm'}><j_{\kappa l}|j_{\kappa' l'}>$$

$$- [V_0\,(r_0,0) - (e^2/\varepsilon r_0)]j_{\kappa l}(r_0)j_{\kappa' l'}(r_0)<Y_{lm}|Y_{l'm'}>\}q_C;$$

in particular:

$$V_{1s\text{-}2s} = \{[V_0(r_0,0)/r_0]<j_{0\kappa}|j_{0\kappa'}> - [V_0\,(r_0,0) - (e^2/\varepsilon r_0)]j_{0\kappa}(r_0)j_{0\kappa'}(r_0)\}q_C = V_{1s\text{-}2s}$$

$$V_{1s\text{-}2pz} = -<Y_{00}|Y_{10}>[V_0(r_0,0) - (e^2/\varepsilon r_0)]j_{\kappa 0}(r_0)j_{\kappa' 1}(r_0)q_C = 0$$

$$V_{2s\text{-}2pz} = -<Y_{00}|Y_{10}>[V_0(r_0,0) - (e^2/\varepsilon r_0)]j_{\kappa 0}(r_0)j_{\kappa' 1}(r_0)q_C = 0,$$

etc., where vanishing results technically from the orthogonality of the angular wave functions at different l. Physically, the off-diagonal terms between different-parity states vanish when mixed by an even-parity vibration, such as the breathing mode.

## Appendix II

### Calculation of the electron-mode coupling constants (The diagonal matrix elements)

The diagonal coupling constants can be derived similar to the off-diagonal constants. This has been done before in Part II of Reference 19 and will now be reproduced as given by equations (4.37) & (4.38) therein. We define $b_{tt} = <t|b(r)|t>$, where $|t>$ is the electronic state, to get:

$$b_{nsns} = (A^2 r_0^3/2v^2)(1/r_0)\{[V_0+\chi+(N/\alpha_M-1)V_M][1 - (1/v)\mathit{sin}v\ \mathit{cos}v]$$

$$- 2(V_0 - e^2/\varepsilon r_0)(\mathit{sin}v)^2\}$$

$$b_{2p2p} = (A^2 r_0^3/2v^2)(1/r_0)\{[V_0+\chi+(N/\alpha_M-1)V_M][1 - (2/v^2)(\mathit{sin}v)^2$$

$$+ (1/v)\mathit{sin}v\ \mathit{cos}v] - 2(V_0 - e^2/\varepsilon r_0)[(1/v)\mathit{sin}v - \mathit{cos}v]^2\}$$

where $V_0$ is the square-well depth and $\varepsilon$ is an appropriate dielectric constant both calculated self-consistently by means of equations (4.3) & (4.6), respectively. $\chi$ is the electron affinity, and $r_0$ is the cavity radius. $\alpha_M$ is Madelung's constant and $V_M = \alpha_M e^2/r_0$ is Madelung's potential.v = $\kappa r_0$ where $\kappa$ is either $\kappa_{ns}$ or $\kappa_{2p}$, as obtained from the pairs of equations (4.17) & (4.24), respectively, while A is either $A_{ns}$ from equation (4.16) or $A_{2p}$ from equation (4.23).

N is the number of neighboring ions to the anion vacancy whose vibration couples to the F center. N = 6 if coupling to the local breathing mode is assumed in a compact electronic ground state and $N = \alpha_M$ for coupling to the $A_{1g}$ lattice mode in an extended electronic excited state.